\documentclass[12pt,draftcls,onecolumn]{IEEEtran}
\usepackage{cite}
\usepackage[font=small]{caption}
\usepackage{ifpdf}
\usepackage{authblk}
\usepackage{commath}
\usepackage{cuted} 
\usepackage{pgfplots}
\usetikzlibrary{arrows,decorations.markings}
\usepackage{bigints}
\newif\ifCLASSOPTIONonecolumn       \CLASSOPTIONonecolumnfalse
\newif\ifCLASSOPTIONtwocolumn       \CLASSOPTIONtwocolumntrue
\usepackage{tikz}
\ifCLASSINFOpdf
\usepackage{graphicx}  
\usepackage{subcaption}
\usepackage{amsmath}
\usepackage{dsfont}
\usepackage{bbold}
\usepackage{amssymb}
\usepackage{algorithmic}
\usepackage{array}
\usepackage{stix}
\usepackage{stfloats}
\fnbelowfloat
\usepackage{bigints}
\begin{document}
%
\title{Performance Evaluation of Adaptive Cooperative NOMA Protocol at Road Junctions}
%
%
%

\author[1]{Baha Eddine Youcef~Belmekki}
\author[2]{Abdelkrim ~Hamza}
\author[1]{Beno\^it~Escrig}
\affil[1]{IRIT Laboratory, School
of ENSEEIHT, Institut National Polytechnique de Toulouse, France,}
\affil[ ]{ e-mail: $\{$bahaeddine.belmekki, benoit.escrig$\}$@enseeiht.fr}
\affil[2]{LISIC Laboratory, Electronic and Computer Faculty, USTHB, Algiers, Algeria,}
\affil[ ]{email: ahamza@usthb.dz}
\setcounter{Maxaffil}{0}
\renewcommand\Affilfont{\small}

\markboth{ }%
{Shell \MakeLowercase{\textit{et al.}}: Bare Demo of IEEEtran.cls for IEEE Journals}

\maketitle

\IEEEpeerreviewmaketitle
 \begin{abstract}
Vehicular communications (VCs) protocols offer useful contributions in the context of accident prevention thanks to the transmission of alert messages. This is even truer at road intersections since these areas exhibit higher collision risks and accidents rate. 
On the other hand, non-orthogonal multiple access (NOMA) has been show to be a suitable candidate for five generation (5G) of wireless systems. 
In this paper, we propose and evaluate the performance of VCs protocol at road intersections, named adaptive cooperative NOMA (ACN) protocol.
The transmission occurs between a source and two destinations. The transmission is subject to interference originated from vehicles located on the roads. The positions of the interfering vehicles follow a Poison point process (PPP).
First, we calculate the outage probability related to ACN protocol, and closed form expressions are obtained. 
Then we compare it with other existing protocols in the literature. We show that ACN protocol offers a significant improvement over the existing protocols in terms of outage probability, especially at the intersection. 
We show that the performance of ACN protocol increases compared to other existing protocols for high data rates.  
The theoretical results are verified with Monte-Carlo simulations. 
\end{abstract}


\section{Introduction}
\subsection{Motivation}
Road traffic safety is a major issue, and more particularly
at road intersections since there areas are more prone to accidents \cite{traficsafety}. In this context, vehicular communications (VCs) protocols provide several contributions for accident prevention thanks to the sending of alert messages. Such applications require high data rates to enable reliable communications. 
To increase data rate and spectral efficiency, non-orthogonal multiple access (NOMA) has been shown to be a suitable candidate for the the fifth generation (5G) of communication systems as a multiple access scheme \cite{ding2014performance}. 
Different from the classical orthogonal multiple access (OMA), NOMA
allows multiple users to share the same resource with different
power allocation levels. Thus, implementing NOMA in VCs will be beneficial when accidents happen and several vehicles have to send alert messages, or informing other vehicles about the accidents status.

On the other hand, cooperative communications have been shown to increase link reliability of wireless networks using two (or more) communication channels with different characteristics, since each channel undergoes different levels of fading and interference \cite{belmekki2019outage}. In this paper, we propose and study the performance of a VCs cooperative NOMA protocol at road junctions.
\subsection{Related Works}
The performance of VCs in the presence of interference have been investigated before. Considering highway scenarios, the authors in \cite{farooq2016stochastic} derivate the expressions for the intensity of concurrent transmitters and packet success probability for multilane highway scenarios considering carrier sense multiple access (CSMA) protocols. The authors in \cite{tong2016stochastic} analyze the performance of IEEE 802.11p using tools from queuing theory and stochastic geometry. The outage probability is obtained in \cite{jiang2016information} for Nakagami-m fading and Rayleigh fading channels. Considering intersection scenarios, the authors in \cite{steinmetz2015stochastic} compute the success probability in the presence of interference considering a direct transmission road intersection scenario. In \cite{abdulla2016vehicle}, the authors calculate the success probability in the presence of interference for intersection scenarios a direct transmission for limited road segments. The performance of vehicle to vehicle (V2V) communications are investigated for multiple intersection streets in \cite{jeyaraj2017reliability}.

As for NOMA, several works investigate the impact of interference in NOMA networks. The authors in \cite{zhang2016stochastic} analyze a downlink NOMA network. In \cite{zhang2017uplink}, the authors analyze a uplink NOMA network. In \cite{zhang2017downlink}, both uplink and downlink are analyzed. The authors of this the paper investigated the impact of NOMA using direct transmission in \cite{J3},  cooperative NOMA at intersections in \cite{J4}, and MRC using NOMA \cite{Belmkki}, and in millimeter wave vehicular communications in \cite{Cmm1,Cmm2}. The authors of this paper also investigated the impact of vehicles mobility, and different transmission schemes on the performance in \cite{J2} and \cite{J1,belmekki2019outagearxiv}, respectively.

Regarding cooperative NOMA protocols, The authors in
 \cite{ding2015cooperative} propose a cooperative NOMA protocol in a half duplex mode with a help of a relay. This conventional cooperative NOMA (CCN) protocol \cite{ding2015cooperative} improves the performance of the transmission by adding a diversity gain. However, the spectral efficiency of this protocol is reduced due to the use of the half duplex mode. 
 To cope with this limitation, the authors in \cite{hu2017efficient} propose a cooperative protocol, named relaying with NOMA back-haul. In  this protocol, the source adjusts the time duration of the transmission based on the global instantaneous channel state information
(CSI). However, global instantaneous CSI at the source can be hard to obtain in practice, especially for real time scenarios such as road safety scenarios. Following this line of research, we propose an adaptive cooperative NOMA (ACN) protocol at road junctions for VCs in the presence of interference.

\subsection{Contributions}
The contributions of this paper are as follows: 
\begin{itemize}
\item We propose and evaluate the performance of VCs protocol at at road intersections in the presence of interference.
\item
We calculate the outage probability related to ACN protocol, and closed form expressions are obtained considering a scenario involving a source, and two destinations.  
\item We compare the performance of ACN protocol with other existing protocols in the literature. We show that ACN protocol offers a significant improvement in terms of outage probability, especially at intersections.
\item We show that the performance of ACN protocol increases compared to other existing protocols for high data rates.
\item All results and the theoretical analysis are verified with Monte Carlo simulations. 

\end{itemize}

\section{System Model}
\begin{figure}[]
\centering
\includegraphics[height=8cm,width=10cm]{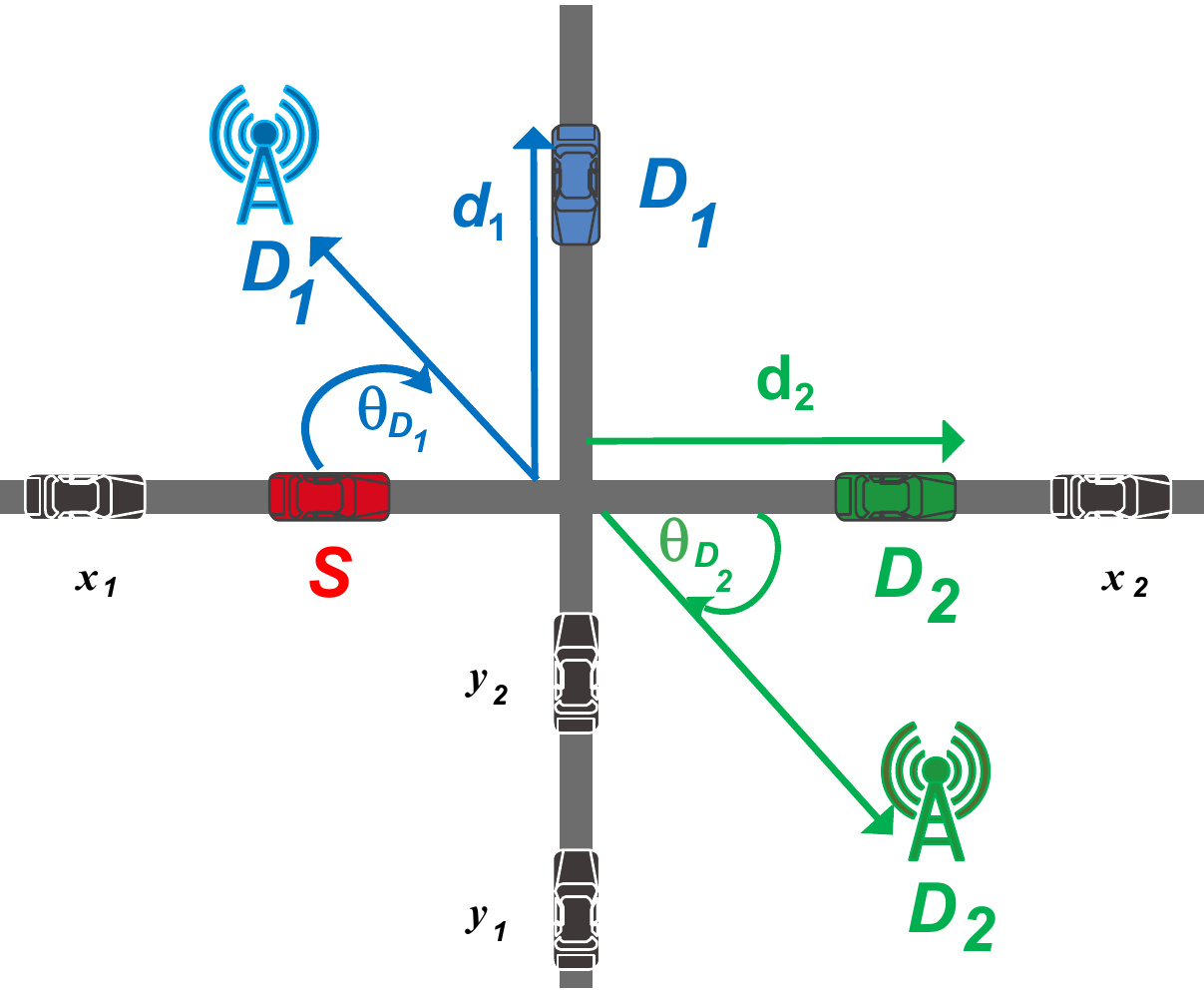}
\caption{NOMA system model for VCs. The nodes $D_1$ and $D_2$ can be vehicles or as part of the communication infrastructure.}
\label{Fig1} 
\end{figure}

In this paper, we consider a NOMA transmission between a source $S$, and two destinations, denoted $D_1$ and $D_2$. The triplet $\{S, D_1, D_2\}$ denotes the nodes and their locations as depicted in Fig.\ref{Fig1}. We consider an intersection scenario involving two perpendicular roads, an horizontal road denoted by $X$, and a vertical road denoted by $Y$.
In this paper, we consider both V2V and vehicle-to-infrastructure (V2I) communications\footnote{The Doppler shift and time-varying effect of V2V and V2I channel are beyond the scope of this paper.}, therefore, any node of the triplet $\lbrace{S, D_1, D_2}\rbrace$ can be either on the road or outside the road. We denote by  $d_i$ and $\theta_i$, the distance between the node $D_i$ and the intersection, 
and the angle between the node $D_i$ and the $X$ road, where $i \in \{1,2\}$ as shown in Fig.\ref{Fig1}. 

The transmission is subject to interference that is originated from vehicles located on the roads. The set of interfering vehicles located on the $X$ road, denoted by $\Phi_{X}$ (resp. on the $Y$ road, denoted by $\Phi_{Y}$) are modeled as a one-dimensional homogeneous Poisson point process (1D-HPPP), i.e, $\Phi_{X}\sim\textrm{1D-HPPP}(\lambda_{X},x)$ (resp.$\Phi_{Y}$ $\sim \textrm{1D-HPPP}(\lambda_{Y},y)$), where $x$ and $\lambda_{X}$ (resp. $y$ and $\lambda_{Y}$) are the position of interfering vehicles and their intensity on the $X$ road (resp. $Y$ road). The notation $x$ and $y$ denotes both the interfering vehicles and their locations. The transmission is subject to path loss between the nodes $a$ and $b$, termed as $l_{ab}$, where $l_{ab}= \Vert a- b\Vert^{-\alpha}$, and $\alpha$ is the path loss exponent. We consider a slotted ALOHA protocol with parameter $p$, i.e., every node can access the medium with a probability $p$ \cite{nguyen2013performance}.

Several works in NOMA order the receiving nodes by their channel states \cite{ding2014performance,ding2015cooperative}.
However, we consider that the receiving nodes are ordered according to their quality of service (QoS) priorities, since it has been show that it is more realistic assumption \cite{ding2016relay,ding2016mimo}. We consider a scenario in which $D_1$ needs low data rate but has to be served immediately, whereas $D_2$ requires high data rate but can be served later. For instance, $D_1$ can be a vehicle that needs to receive safety data information about an accident in its surrounding, whereas $D_2$ can be a user that accesses an internet connection. We also consider an interference limited scenario, and thus, we set the power of the additive noise to zero. We assume, without loss of generality, that all nodes transmit with a unit power. The signal transmitted by $S$, denoted $ \chi_{S}$, is a mixture of the message intended to $D_1$ and $D_2$. This can be expressed as
\begin{equation}
 \chi_{S}=\sqrt{a_1}\chi_{D_1}+\sqrt{a_2}\chi_{D_2}, \nonumber
 \end{equation}
where $a_i$ is the power coefficients allocated to $D_i$, and $\chi_{D_i}$ is the message intended to $D_i$. Since $D_1$ has a higher power allocation than $D_2$, that is, $a_1 \ge a_2$, then $D_1$ comes first in the decoding order. Note that, $a_1+a_2=1$.\\
 The signal received at $D_i$ is then expressed as
 \begin{equation}
   \mathcal{Y}_{D_i}=h_{SD_i}\sqrt{l_{SD_i}}\:\chi_{S}+ 
 \sum_{x\in \Phi_{X_{D_i}}}h_{D_ix}\sqrt{l_{D_ix}}\:\chi_{x} 
 +\sum_{y\in \Phi_{Y_{D_i}}}h_{D_iy}\sqrt{l_{D_iy}}\:\chi_{y}, \nonumber
 \end{equation}
where $\mathcal{Y}_{D_i}$ is the signal received by $D_i$.
The messages transmitted by the interfering node $x$ and $y$, are denoted respectively by $ \chi_x$ and $\chi_y $, $h_{ab}$ denotes the fading coefficient between node $a$ and $b$, and it is modeled as $\mathcal{CN}(0,1)$. The power fading coefficient between the node $a$ and $b$, denoted $|h_{ab}|^2$, follows an exponential distribution with unit mean.
The aggregate interference is defined as 
\begin{eqnarray}
I_{X_{D_i}}=\sum_{x\in \Phi_{X_{D_i}}}\vert h_{D_ix}\vert^{2}l_{D_ix},   \\ 
I_{Y_{D_i}}=\sum_{y\in \Phi_{Y_{D_i}}}\vert h_{D_iy}\vert^{2}l_{D_iy} , 
\end{eqnarray}
where $I_{X_{D_i}} $ denotes the aggregate interference from the $X$ road at $D_i$, $I_{Y_{D_i}}$ denotes the aggregate interference from the $Y$road at $D_i$, $\Phi_{X_{D_i}}$ denotes the set of the interferers from the $X$ road at $D_i$, and $\Phi_{Y_{D_i}}$ denotes the set of the interferers from the $Y$ road at $D_i$.

\section{ACN Protocol}
First, we consider the scenario in which  $D_1$ acts as relay to transmit the message to $D_2$ \footnote{The relay selection algorithms are out of the scope of this paper.}.
At the beginning of each transmission, $S$ sends the superimposed signal
to $D_1$ and $D_2$ using a direct transmission \cite{C1}. 
If $D_2$ decodes its desired message, it sends a 1-bit positive acknowledgement (ACK) to $S$ and $D_1$, and thus, the transmission occurs in one phase.
However, if $D_2$ is unable to decode its desired message, it sends a 1-bit negative acknowledge (NACK) to $S$ and $D_1$. Hence, if $D_1$ decodes its desired message and $D_2$ message, it sends $D_2$ message using cooperative transmission \cite{belmekki2019outage} using OMA. Thus, the transmission occurs in two phases.

\begin{figure}[]
\centering
\includegraphics[scale=0.8]{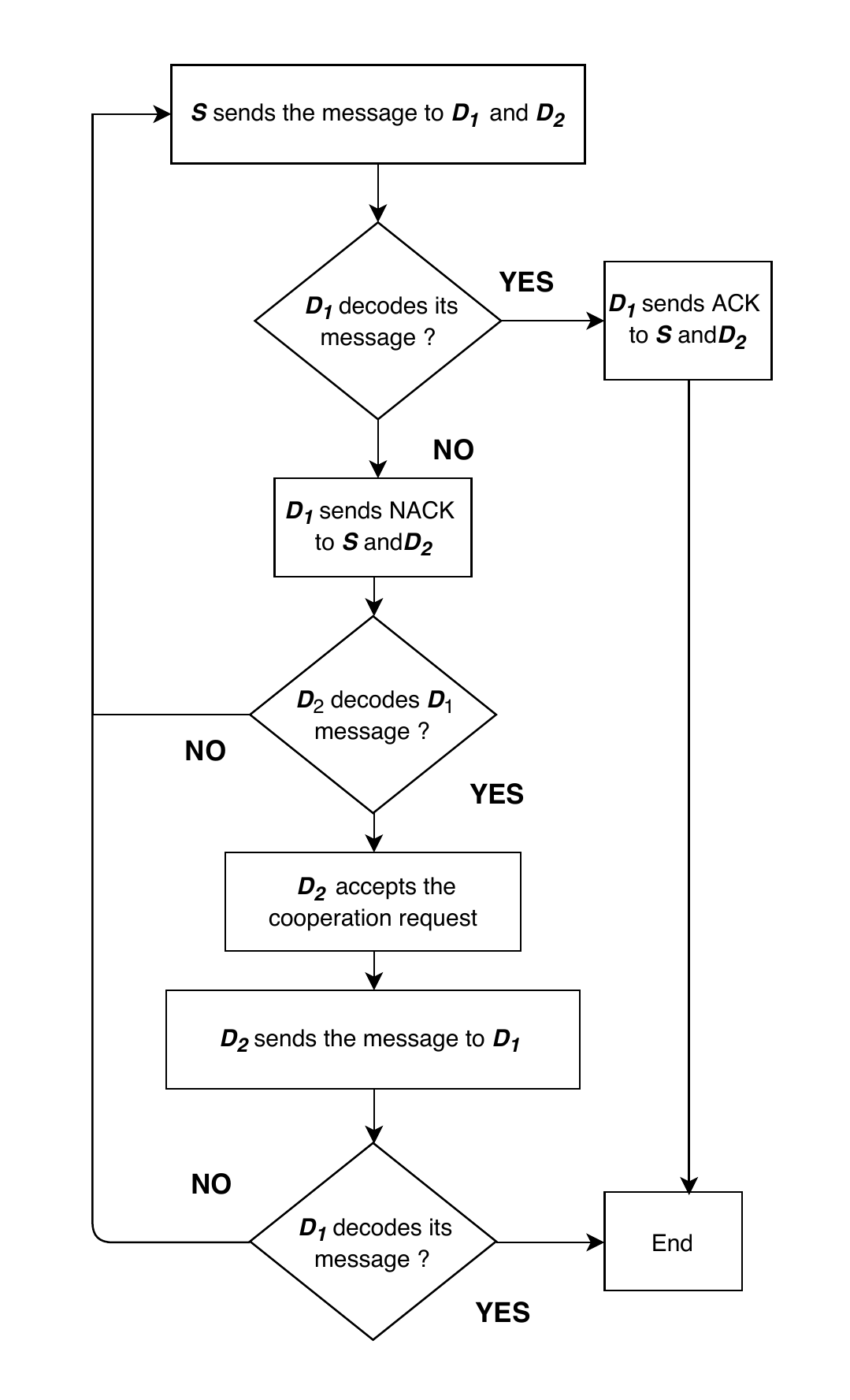}
\caption{Flow chart of ACN protocol at $D_1$.}
\label{Fig2}
\end{figure}

\begin{figure}[]
\centering
\includegraphics[scale=0.8]{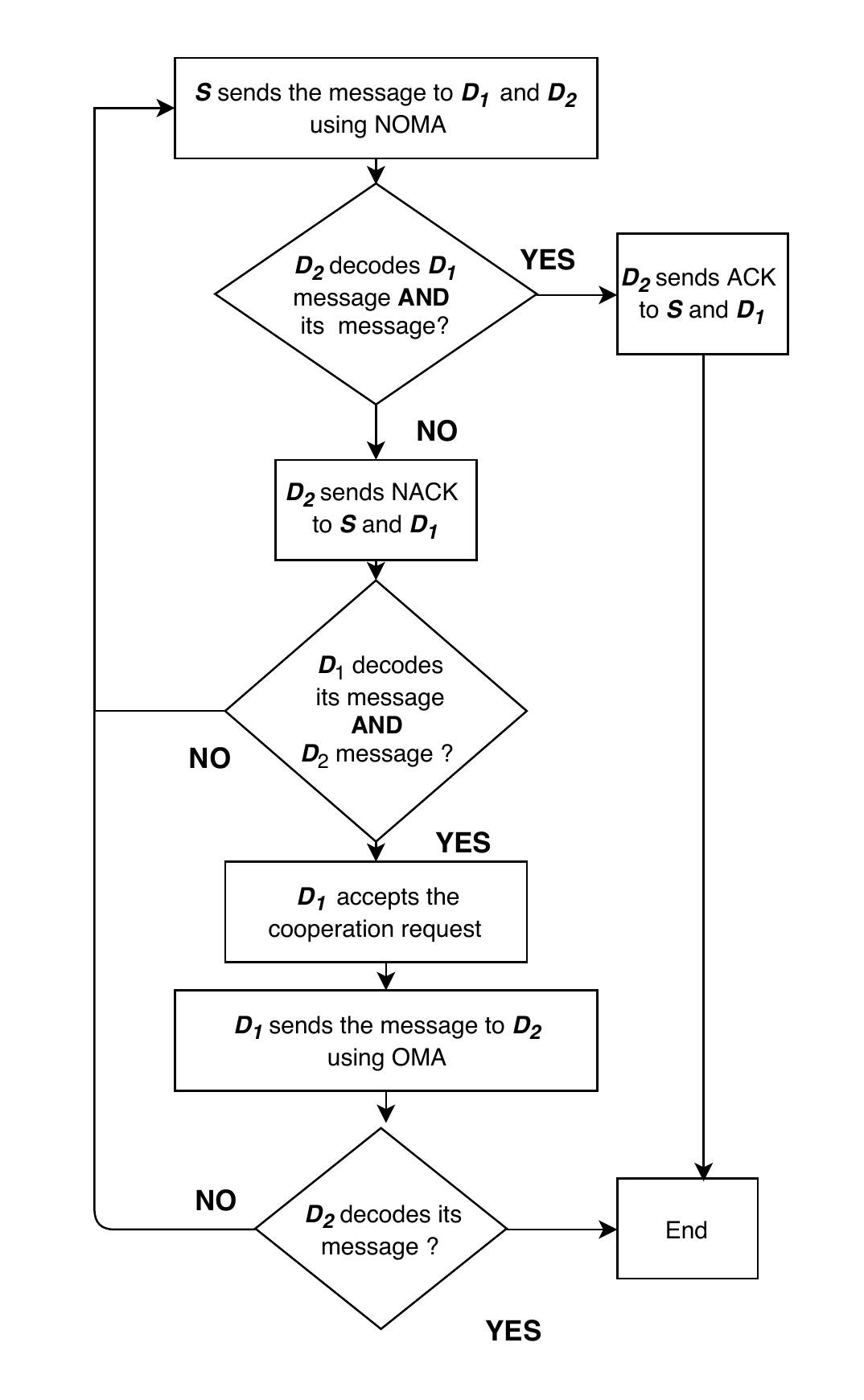}
\caption{Flow chart of ACN protocol at $D_2$.}
\label{Fig3}
\end{figure}

Now, we consider the scenario in which $D_2$ acts as relay to transmit the message to $D_1$.
In this same way, $S$ sends the superimposed message
to $D_1$ and $D_2$ using a direct transmission. 
If $D_1$ decodes its desired message, it sends a 1-bit ACK to $S$ and $D_2$, and thus, the transmission occurs in one phase.
However, if $D_1$ is unable to decode its desired message, it sends a 1-bit NACK to $S$ and $D_1$. Hence, if $D_2$ decodes $D_1$ message, it sends $D_1$ message using cooperative transmission, and without using NOMA. Thus, the transmission occurs in two phases\footnote{Note that ACN protocol does no need to perform
channel estimation to switch between direct transmission and cooperative transmission.}. The flow charts of ACN protocol related to $D_1$ and $D_2$ are respectively given by Fig.\ref{Fig2} and Fig.\ref{Fig3}. The ACN protocol switches to cooperative transmission only if the direct transmission is not feasible. This will induce a latency because the transmission will occur during two time slots instead of one time slot. However, as we will show in Section \ref{results}, the ACN protocol increases the performance in terms of outage probability compared to other transmission schemes and protocols in the literature.

\section{ACN Protocol Outage Expressions}
\subsection{Signal-to-Interference Ratio (SIR) Expressions}

The outage probability is defined as the probability that the signal-to-interference ratio (SIR) at the receiver node is below a given threshold. According to successive interference cancellation (SIC) \cite{hasna2003performance}, $D_1$ will be decoded first since it has the higher power allocation, and $D_2$ message will be considered as interference. The SIR at $D_1$ to decode its desired message, denoted $\textrm{SIR}_{D_{1-1}}$, is expressed as
\begin{equation}\label{eqq}
\textrm{SIR}_{D_{1-1}}=\frac{\vert h_{SD_1}\vert^{2}l_{SD_1}\,a_1}{\vert h_{SD_1}\vert^{2}l_{SD_1}a_2+I_{X_{D_1}}+I_{Y_{D_1}}} .
\end{equation}
Similarly, The SIR at $D_1$ to decode $D_2$ message, denoted $\textrm{SIR}_{D_{1-2}}$, is expressed as\footnote{Perfect SIC is considered in this work, that is, no fraction of power remains after SIC process.}
\begin{equation}
\textrm{SIR}_{D_{1-2}}=\frac{\vert h_{SD_1}\vert^{2}l_{SD_1}\,a_2}{I_{X_{D_1}}+I_{Y_{D_1}}} .
\end{equation}
Since $D_2$ has the lower power allocation, it decodes $D_1$ message first, then decodes its intended message. The SIR at $D_2$ to decode $D_1$ message, denoted $\textrm{SIR}_{D_{2-1}}$, is expressed as
\begin{equation}
\textrm{SIR}_{D_{2-1}}=\frac{\vert h_{SD_2}\vert^{2}l_{SD_2}\,a_1}{\vert h_{SD_2}\vert^{2}l_{SD_2}a_2+I_{X_{D_2}}+I_{Y_{D_2}}}.
\end{equation}
The SIR at $D_2$ to decode its desired message, denoted $\textrm{SIR}_{D_{2-2}}$, is expressed as  
\begin{equation}
\textrm{SIR}_{D_{2-2}}=\frac{\vert h_{SD_2}\vert^{2}l_{SD_2}\,a_2}{I_{X_{D_2}}+I_{Y_{D_2}}}.
\end{equation}
When using the cooperative transmission, the node that acts as a relay uses OMA instead of NOMA, since the transmission involves only one receiving node. Hence, the SIR at the receiver is then expressed as 
\begin{equation}
\textrm{SIR}^{(\textrm{OMA})}_{D_{k}D_{l}}=\frac{\vert h_{D_kD_l}\vert^{2}l_{D_{k}D_{l}}}{I_{X_{D_l}}+I_{Y_{D_l}}},
\end{equation}
where $\{k,l\}\in\{1,2\}$.

\subsection{ACN Outage Event Expressions}
Now, we will express the outage events related to the ACN protocol for $D_1$ and $D_2$. The outage events related to $D_1$ and $D_2$ using ACN protocol, denoted respectively by $\mathcal{O}_{\textrm{ACN}}(D_1)$ and $\mathcal{O}_{\textrm{ACN}}(D_2)$, can be expressed as 
\begin{equation}
\mathcal{O}_{\textrm{ACN}}(D_1)
= 1-\mathcal{O}^C_{\textrm{ACN}}(D_1),
\end{equation}
and
\begin{equation}
\mathcal{O}_{\textrm{ACN}}(D_2)
= 1-\mathcal{O}^C_{\textrm{ACN}}(D_2),
\end{equation}
where $\mathcal{O}_{\textrm{ACN}}(D_1)$ and $\mathcal{O}_{\textrm{ACN}}(D_2)$ denote respectively the success events related to $D_1$ and $D_2$.
The expression of $\mathcal{O}^C_{\textrm{ACN}}(D_1)$ and $\mathcal{O}^C_{\textrm{ACN}}(D_2)$ are respectively given by 
\begin{equation}
\mathcal{O}^C_{\textrm{ACN}}(D_1)
= \{\textrm{DT}^C_{SD_1}\}  \cup \{ \textrm{DT}_{SD_1} \cap \textrm{RT}^C_{S,D_2,D_1} \},
\end{equation}
and
\begin{equation}
\mathcal{O}^C_{\textrm{ACN}}(D_2)
= \{\textrm{DT}^C_{SD_2}\}  \cup \{ \textrm{DT}_{SD_2} \cap \textrm{RT}^C_{S,D_1,D_2} \},
\end{equation}
where $\textrm{DT}^C_{SD_n}$, $\textrm{RT}^C_{S,D_2,D_1}$, and $\textrm{RT}^C_{S,D_1,D_2}$, are expressed as
\begin{equation}
\textrm{DT}^C_{SD_n}= \bigcap^{n}_{i=1} \textrm{SIR}_{SD_{n-i}}<\Theta^{(1)}_{i}, 
\end{equation}
and
\begin{equation}
\textrm{RT}^C_{S,D_2,D_1}=\Big\{\textrm{SIR}_{SD_{2-1}}\ge\Theta^{(2)}_{1} \cap \textrm{SIR}^{(\textrm{OMA})}_{D_{2}D_{1}} \ge \Theta^{(2)}_{1}\Big\},
\end{equation}
and
\begin{equation}
\textrm{RT}^C_{S,D_1,D_2}=\Big\{\bigcap^{2}_{i=1} \textrm{SIR}_{SD_{1-i}}\ge\Theta^{(2)}_{i} \cap \textrm{SIR}^{(\textrm{OMA})}_{D_{1}D_{2}} \ge\Theta^{(2)}_{2}\Big\}.
\end{equation}
The decoding threshold $\Theta^{(n)}_{i}$ is defined as
\begin{equation}
\Theta^{(n)}_{i}\triangleq2^{n\mathcal{R}_i}-1, 
\end{equation}
where $\mathcal{R}_i$ is the target data rate of $D_i$. Note that, $n=1$ when direct transmission is used, and $n=2$ when cooperative transmission is used.
\subsection{ACN Outage Probability Expressions}

In the following, we will express the probabilities related to $\mathcal{O}_{\textrm{ACN}}(D_1)$ and $\mathcal{O}_{\textrm{ACN}}(D_2)$. The outage probability expressions related to $D_1$ and $D_2$, denoted $\mathbb{P}\big[\mathcal{O}_{\textrm{ACN}}(D_1)\big]$ and $\mathbb{P}\big[\mathcal{O}_{\textrm{ACN}}(D_2)\big]$, are respectively given by

\begin{equation}\label{eq.18}
   \mathbb{P}\big[\mathcal{O}_{\textrm{ACN}}(D_1)\big]=1-\left[\mathcal{W}_{(D_1)}\Big(\frac{\mathcal{G}^{(1)}_{1}}{l_{SD_1}}\Big)+\Bigg\{   \Bigg(1-\mathcal{W}_{(D_1)}\Big(\frac{\mathcal{G}^{(1)}_{1}}{l_{SD_1}}\Big)\Bigg) \times 
   \mathcal{W}_{(D_2)}\Big(\frac{\mathcal{G}^{(2)}_{1}}{l_{SD_2}}\Big) \times 
    \mathcal{W}_{(D_1)}\Big(\frac{{\Theta}^{(2)}_{1}}{l_{D_2D_1}}\Big)  \Bigg\}\right],
\end{equation}
and
\begin{equation}\label{eq.19}
   \mathbb{P}\big[\mathcal{O}_{\textrm{ACN}}(D_2)\big]=1-\left[\mathcal{W}_{(D_2)}\Big(\frac{\mathcal{G}^{(1)}_{\max}}{l_{SD_2}}\Big)+\Bigg\{   \Bigg(1-\mathcal{W}_{(D_2)}\Big(\frac{\mathcal{G}^{(1)}_{\max}}{l_{SD_2}}\Big)\Bigg) \times 
   \mathcal{W}_{(D_1)}\Big(\frac{\mathcal{G}^{(2)}_{\max}}{l_{SD_1}}\Big) \times 
    \mathcal{W}_{(D_2)}\Big(\frac{{\Theta}^{(2)}_{2}}{l_{D_1D_2}}\Big)  \Bigg\}\right],
\end{equation}
where the function $\mathcal{W}_{(D_i)}\Big(\frac{A}{B}\Big)$ is given by 
\begin{equation}\label{eq.20}
    \mathcal{W}_{(D_i)}\Big(\frac{A}{B}\Big)=\mathcal{L}_{I_{X_{D_i}}}\Big(\frac{A}{B}\Big)\mathcal{L}_{I_{Y_{D_i}}}\Big(\frac{A}{B}\Big),
\end{equation}
and $\mathcal{G}^{(n)}_{1}$ and $\mathcal{G}^{(n)}_{\mathrm{max}}$, are respectively given by
\begin{equation}
 \mathcal{G}^{(n)}_{1}=\frac{\Theta^{(n)}_1}{a_1-\Theta^{(n)}_1 a_2},  
\end{equation}
and
\begin{equation}
 \mathcal{G}^{(n)}_{\mathrm{max}}=\mathrm{max}(\mathcal{G}^{(n)}_{1},\mathcal{G}^{(n)}_{2}),  
\end{equation}
where $\mathcal{G}^{(n)}_{2}=\Theta^{(n)}_2 /a_2$.\\
\textit{Proof}:  See Appendix A.\hfill $ \blacksquare $ \\

\begin{figure}[]
\centering
\includegraphics[scale=0.6]{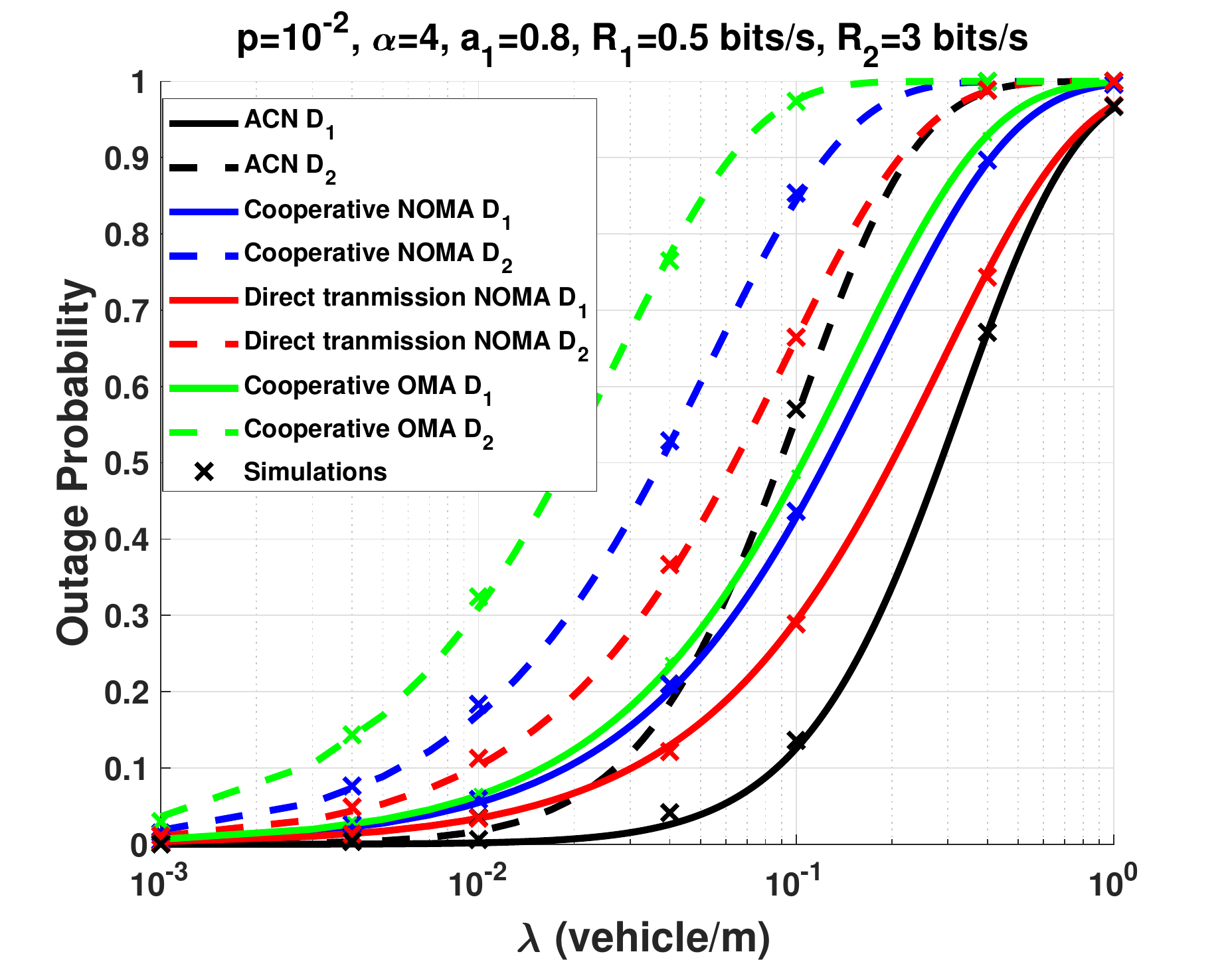}
\caption{Outage probability as a function of $\lambda$ considering ACN, cooperative NOMA, direct transmission NOMA, and cooperative OMA.}
\label{Fig4}
\end{figure}
The Laplace transform expressions, $\mathcal{L}_{I_{X_{Di}}}$ and $\mathcal{L}_{I_{Y_{Di}}}$, are respectively given by
\begin{equation}\label{355}
\mathcal{L}_{I_{X_{Di}}}(s)=\exp\Bigg(\dfrac{-s\:\emph{p}\lambda_{X}\pi}{\sqrt{s+d_{i}^2\sin(\theta_{{D_i}})^2}}\Bigg),
\end{equation}
and
\begin{equation}\label{366}
\mathcal{L}_{I_{Y_{Di}}}(s)=\exp\Bigg(\dfrac{-s\:\emph{p}\lambda_{Y}\pi}
{\sqrt{ s+d_{i}^2\cos(\theta_{{D_i}})^2}}\Bigg).
\end{equation}
\textit{Proof}:  See Appendix C in \cite{belmekki2019outage}.\hfill $ \blacksquare $ \\

\section{Simulations and Discussions}\label{results}

\begin{figure}[]
\centering
\includegraphics[scale=0.6]{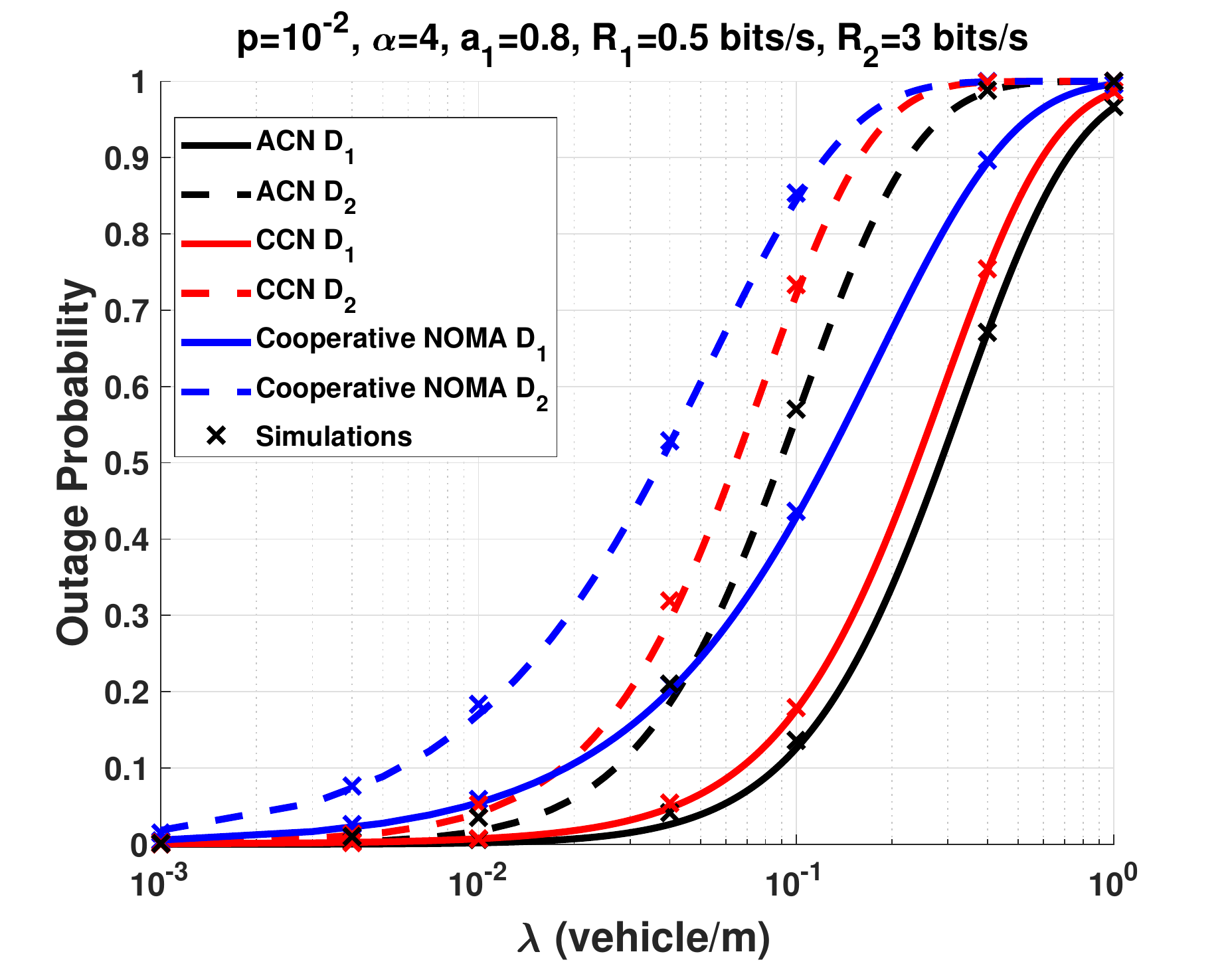}
\caption{Outage probability as a function of $\lambda$ considering ACN, CCN, and cooperative NOMA.}
\label{Fig5}
\end{figure}

In this section, we evaluate the performance of NOMA at road intersections. To verify the accuracy of our theoretical analysis, Monte Carlo simulations are performed by averaging over 50,000 realizations of PPPs and fading channel parameters. In all figures, the marks represent the Monte Carlo simulations \footnote{The confidence intervals in the simulations are very small, this is why they have been omitted.}. We set, without loss of generality,  $\lambda_X = \lambda_Y = \lambda$. \\

Fig.\ref{Fig4} shows the outage probability as a function of $\lambda$ considering ACN, cooperative transmission using NOMA \cite{belmekki2019outage}, direct transmission using NOMA \cite{Belm1904:Outage}, and the classical cooperative OMA.
We can see from Fig.\ref{Fig4}, that as the intensity of vehicles $\lambda$ increases, the outage probability increases. This is because as the intensity increases, the number of interfering vehicles increases, which decreases the SIR at the receiving node.
We can also see from Fig.\ref{Fig4}, that the ACN protocol outperforms the cooperative transmission using NOMA, direct transmission using NOMA, and the classical cooperative OMA. This is because, the ACN protocol can switch its transmission scheme. Hence, it uses the direct transmission in the first phase, when it fails,
it switch to the cooperative transmission in the second phase. 

\begin{figure}[]
\centering
\includegraphics[scale=0.6]{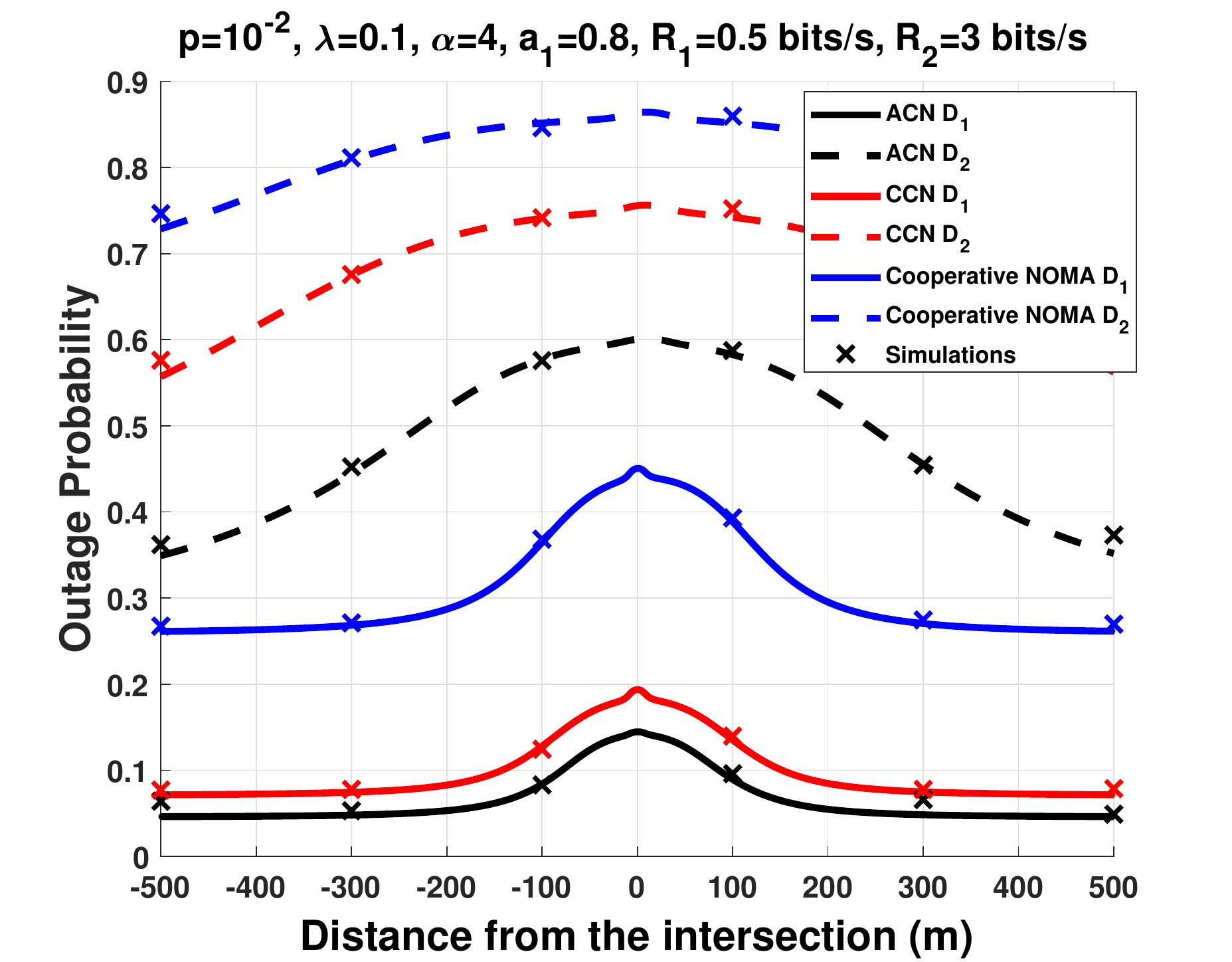}
\caption{Outage probability as a function of the distance from the intersection, considering ACN, CCN, and cooperative NOMA.}
\label{Fig6}
\end{figure}

\begin{figure}[]
\centering
\includegraphics[scale=0.6]{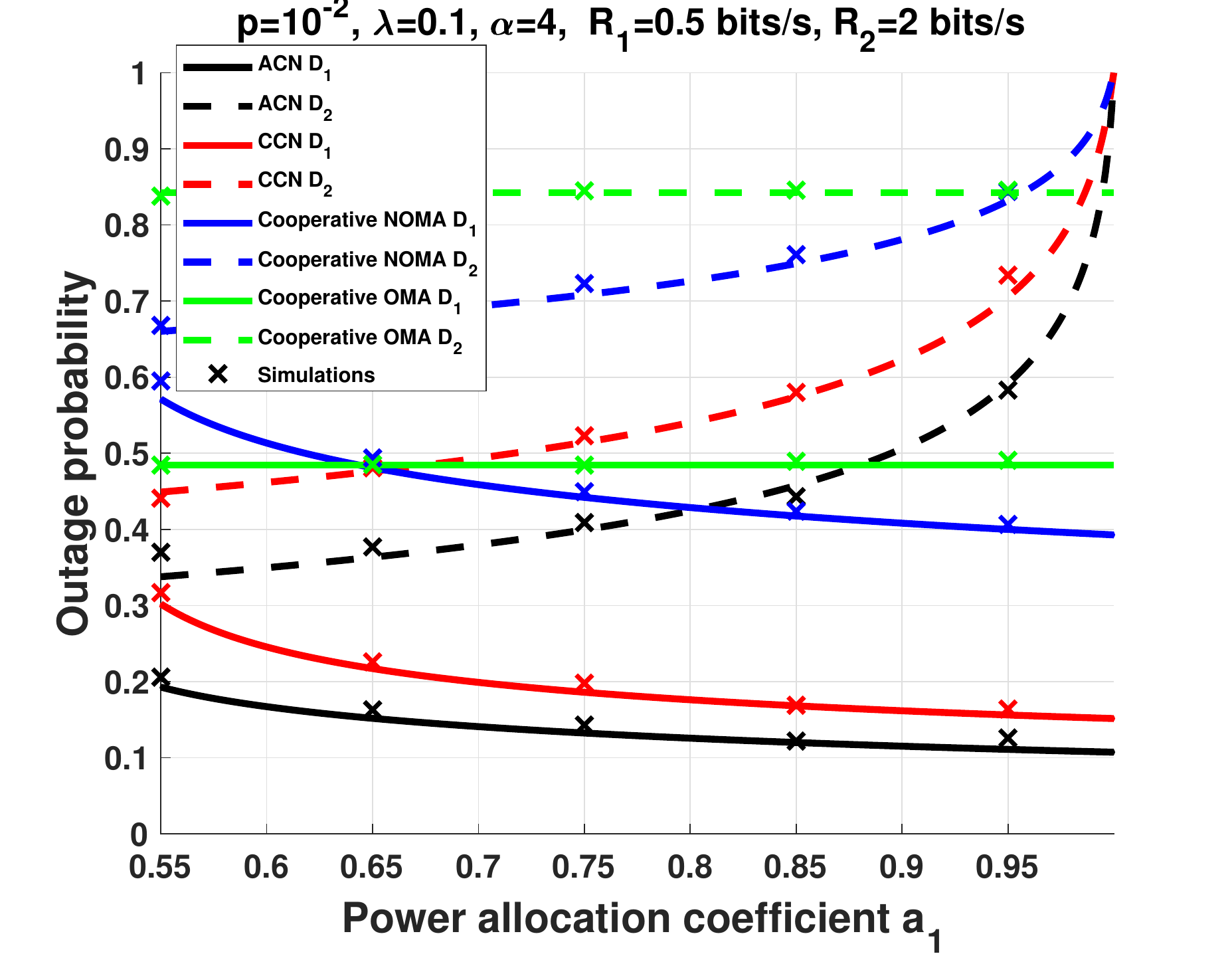}
\caption{Outage probability as a function of the power allocation coefficient $a_1$, considering ACN, CCN, cooperative NOMA, and cooperative OMA.}
\label{Fig7}
\end{figure}

Fig.\ref{Fig5} shows the outage probability as a function of $\lambda$ considering ACN, CCN \cite{ding2015cooperative}, and cooperative NOMA.
We can see from Fig.\ref{Fig5}, that both ACN and CCN outperform the cooperative transmission using NOMA. We can also see that ACN outperforms CCN for both $D_1$ and $D_2$. This is because the transmission in CCN occurs in two phases, hence it reduces its spectral efficiency. On the other hand, the ACN protocol occurs in one phase if the direct transmission succeed, which increases the spectral efficiency compared to CCN. Also, during the second phase of the cooperative transmission, the ACN use OMA to transmit the message since there is only one signal to transmit in the second hop ($D_1 \rightarrow D_2$ or $D_2 \rightarrow  D_1$). Hence, it increases the SIR at the receiver node.

Fig.\ref{Fig6} plots the outage probability as a function of the nodes distance from the intersection, considering ACN, CCN, and cooperative NOMA. We assume that the distance between $S$, $D_1$, and $D_2$ does not change through the simulation. Hence, the nodes of triplet $\{S,D_1,D_2\}$ move together towards the intersection. We set $\Vert S-D_1\Vert=\Vert S-D_2 \Vert=100$ m.
We can see from Fig.\ref{Fig6}, that as the nodes come closer to the intersection (200 m for $D_1$ and 500 m for $D_2$), the outage probability increases. This is because, when the nodes are at the intersection, the interfering vehicles form the $X$ road and the $Y$ road both contribute to the aggregate interference, which decreases the SIR at the receiving nodes.
We can also see that, ACN outperforms both CCN and cooperative NOMA at the intersection. However, we can see that there is a big gap in performance between ACN and CCN regarding $D_2$. This is because, the spectral efficiency of CCN decrease drastically for high data rates. This is why, ACN protocol offers a better performance for high data rates compared to CCN. 
Finally, we can see in CCN and cooperative NOMA, that the outage probability increases more in the last $10$ m. However, there is no increases in the outage probability when using ACN.

Fig.\ref{Fig7} plots the outage probability as a function pf $a_1$, considering ACN, CCN, cooperative  NOMA, and cooperative OMA. We can see from Fig.\ref{Fig7}, that ACN outperforms CCN and cooperative NOMA and cooperative OMA regardless of $a_1$ value. We can also see that when $a_1$ increases, the outage probability of $D_1$ decreases, whereas the outage probability of $D_2$ decreases. 
Finally, we can see that the performance of ACN are greater for $D_2$, since $D_2$ has a high data rate.

\section{Conclusion}

In this paper, we proposed and evaluated the performance of ACN protocol at road intersections.
We calculated the outage probability related to ACN protocol, and closed form expressions were obtained for two destinations nodes. 
We compared the ACN protocol with cooperative NOMA protocol, direct NOMA protocol, and the classical cooperative OMA protocol, and we showed that ACN protocol outperforms these protocols in terms of outage probability, especially at intersections. 
We also compared the performance of ACN protocol with the CCN protocol, and we showed that the ACN protocol offers better performance than CCN protocol at road intersections in terms of outage probability.
Finally, we showed that the performance of ACN protocol increases compared to other existing protocols for high data rates.  
 
\appendices
\section{}
The probability $\mathbb{P}\big[\mathcal{O}_{\textrm{ACN}}(D_1)\big]$ is expressed as 
\begin{equation}
\mathbb{P}\big[\mathcal{O}_{\textrm{ACN}}(D_1)\big]
= 1-\mathbb{P}\big[\mathcal{O}^C_{\textrm{ACN}}(D_1)\big].
\end{equation}
The probability $\mathbb{P}\big[\mathcal{O}^C_{\textrm{ACN}}(D_1)\big]$ is given by
\begin{equation}
\mathbb{P}\big[\mathcal{O}^C_{\textrm{ACN}}(D_1)\big]
=   
\mathbb{P}\left(\textrm{DT}^C_{SD_1}\right)+ \left\{\mathbb{P}\left(\textrm{DT}_{SD_1}\right) \times \mathbb{P}\left(\textrm{RT}^C_{S,D_2,D_1}\right)\right\}.
\end{equation}
To calculate $\mathbb{P}\left(\textrm{DT}^C_{SD_1}\right)$, we proceed as follows 
\begin{equation}\label{eqa1}
\mathbb{P}\left(\textrm{DT}^C_{SD_1}\right)=\mathbb{P}\left(\textrm{SIR}_{SD_{1-1}}\ge \Theta^{(1)}_{1}\right)
\end{equation}
plugging (\ref{eqq}) into (\ref{eqa1}), we get
\begin{align}\label{eqa2}
\mathbb{P}\left(\textrm{DT}^C_{SD_1}\right)=& \mathbb{E}_{I_{X},I_Y}\Bigg[\mathbb{P}\Bigg\lbrace\frac{\vert h_{SD_1}\vert^{2}l_{SD_1}a_1}{\vert h_{SD_1}\vert^{2}l_{SD_1}a_2+I_{X_{D_1}}+I_{Y_{D_1}}} \ge \Theta^{(1)}_{1}\Bigg\rbrace\Bigg]\nonumber\\ 
=&\mathbb{E}_{I_{X},I_Y}\Bigg[\mathbb{P}\Bigg\lbrace\vert h_{SD_1}\vert^{2}l_{SD_1}(a_1-\Theta^{(1)}_{1} a_2)\ge  \Theta^{(1)}_{1}\big[I_{X_{D_1}}+I_{Y_{D_1}}\big]\Bigg\rbrace\Bigg]. 
\end{align}
We can see from (\ref{eqa2}) that, when $\Theta^{(1)}_{1} \ge a_1/ a_2$, the success probability $\mathbb{P}\left(\textrm{DT}^C_{SD_1}\right)=0$ . Then, when $\Theta^{(1)}_{1} < a_1/ a_2$, and after setting $\mathcal{G}^{(1)}_{1}= \Theta^{(1)}_{1} /(a_1- \Theta^{(1)}_{1} a_2)$, we get
\begin{equation}
\mathbb{P}\left(\textrm{DT}^C_{SD_1}\right) =\mathbb{E}_{I_{X},I_Y}\Bigg[\mathbb{P}\Bigg\lbrace\vert h_{SD_1}\vert^{2}\ge\frac{\mathcal{G}^{(1)}_{1}}{l_{SD_1}}\big[I_{X_{D_1}}+I_{Y_{D_1}}\big]\Bigg\rbrace\Bigg].\nonumber
\end{equation}
Since $|h_{SD_1}|^2$ follows an exponential distribution with unit mean, and using the independence of the PPP on the road $X$ and $Y$, we obtain
\begin{equation}
\mathbb{P}\left(\textrm{DT}^C_{SD_1}\right) =\mathbb{E}_{I_{X}}\Bigg[\exp\Bigg(-\frac{\mathcal{G}^{(1)}_{1}}{l_{SD_1}}I_{X_{D_1}}\Bigg)\Bigg]\mathbb{E}_{I_Y}\Bigg[\exp\Bigg(-\frac{\mathcal{G}^{(1)}_{1}}{l_{SD_1}}I_{Y_{D_1}}\Bigg)\Bigg].\nonumber
\end{equation}
Given that  $\mathbb{E}[e^{sI}]=\mathcal{L}_I(s)$,  we finally get
\begin{equation}
\mathbb{P}\left(\textrm{DT}^C_{SD_1}\right)\\ =\mathcal{L}_{I_{X_{D_1}}}\bigg(\frac{\mathcal{G}^{(1)}_{1}}{l_{SD_1}}\bigg)\mathcal{L}_{I_{Y_{D_1}}}\bigg(\frac{\mathcal{G}^{(1)}_{1}}{l_{SD_1}}\bigg).
\end{equation}
Following the same steps, we obtain
\begin{equation}
\mathbb{P}\left(\textrm{DT}_{SD_1}\right)\\ =1-\mathcal{L}_{I_{X_{D_1}}}\bigg(\frac{\mathcal{G}^{(1)}_{1}}{l_{SD_1}}\bigg)\mathcal{L}_{I_{Y_{D_1}}}\bigg(\frac{\mathcal{G}^{(1)}_{1}}{l_{SD_1}}\bigg).
\end{equation}
To calculate $\mathbb{P}\left(\textrm{RT}^C_{S,D_2,D_1}\right)$, we proceed as follows 
\begin{equation}
 \mathbb{P}\left(\textrm{RT}^C_{S,D_2,D_1}\right)=  \mathbb{P}\left(\textrm{SIR}_{SD_{2-1}}\ge\Theta^{(2)}_{1}\right) \times \mathbb{P}\left(\textrm{SIR}^{(\textrm{OMA})}_{D_{2}D_{1}} \ge \Theta^{(2)}_{1} \right).  
\end{equation}
The probability $\mathbb{P}\left(\textrm{SIR}_{SD_{2-1}}\ge\Theta^{(2)}_{1}\right)$ can be acquired following the same steps above, and it is given by
\begin{equation}
\mathbb{P}\left(\textrm{SIR}_{SD_{2-1}}\ge\Theta^{(2)}_{1}\right)=\mathcal{L}_{I_{X_{D_2}}}\bigg(\frac{\mathcal{G}^{(2)}_{1}}{l_{SD_2}}\bigg)\mathcal{L}_{I_{Y_{D_2}}}\bigg(\frac{\mathcal{G}^{(2)}_{1}}{l_{SD_2}}\bigg).
\end{equation}
The probability $\mathbb{P}\left(\textrm{SIR}^{(\textrm{OMA})}_{D_{2}D_{1}} \ge \Theta^{(2)}_{1} \right)$ can be easily calculated, and it is given by
\begin{equation}
\mathbb{P}\left(\textrm{SIR}^{(\textrm{OMA})}_{D_{2}D_{1}} \ge \Theta^{(2)}_{1} \right)=    
\mathcal{L}_{I_{X_{D_1}}}\bigg(\frac{\Theta^{(2)}_{1}}{l_{D_{2}D_1}}\bigg)\mathcal{L}_{I_{Y_{D_1}}}\bigg(\frac{\Theta^{(2)}_{1}}{l_{D_{2}D_1}}\bigg).
\end{equation}

In the same way, we express The probability $\mathbb{P}\big[\mathcal{O}_{\textrm{ACN}}(D_2)\big]$  as a function of a success probability $\mathbb{P}\big[\mathcal{O}^C_{\textrm{ACN}}(D_2)\big]$ as follows 
\begin{equation}
\mathbb{P}\big[\mathcal{O}_{\textrm{ACN}}(D_2)\big]
= 1-\mathbb{P}\big[\mathcal{O}^C_{\textrm{ACN}}(D_2)\big].
\end{equation}
The probability $\mathbb{P}\big[\mathcal{O}^C_{\textrm{ACN}}(D_2)\big]$ is given by
\begin{equation}
\mathbb{P}\big[\mathcal{O}^C_{\textrm{ACN}}(D_2)\big]
=   
\mathbb{P}\left(\textrm{DT}^C_{SD_2}\right)+ \left\{\mathbb{P}\left(\textrm{DT}_{SD_2}\right) \times \mathbb{P}\left(\textrm{RT}^C_{S,D_1,D_2}\right)\right\}.
\end{equation}
To calculate $\mathbb{P}\left(\textrm{DT}^C_{SD_2}\right)$, we proceed as follows 
\begin{align}
\mathbb{P}\left(\textrm{DT}^C_{SD_2}\right)=&\mathbb{P}\left(\bigcap^{2}_{i=1}\textrm{SIR}_{SD_{2-i}}\ge \Theta^{(1)}_{i}\right)\nonumber\\
=&\mathbb{P}\left(\textrm{SIR}_{SD_{2-1}}\ge \Theta^{(1)}_{1} \cap \textrm{SIR}_{SD_{2-2}}\ge \Theta^{(1)}_{2}\right).
\end{align}

Following the same steps as for $\mathbb{P}\left(\textrm{DT}^C_{SD_1}\right)$, we get
\begin{equation}
\mathbb{P}\left(\textrm{DT}^C_{SD_2}\right)= \mathbb{E}_{I_{X},I_Y}\Bigg[\mathbb{P}\Bigg\lbrace\frac{\vert h_{SD_2}\vert^{2}l_{SD_2}a_1}{\vert h_{SD_2}\vert^{2}l_{SD_2}a_2+I_{X_{D_2}}+I_{Y_{D_2}}} \ge \Theta^{(1)}_{1} , \frac{\vert h_{SD_2}\vert^{2}l_{SD_2}a_2}{I_{X_{D_2}}+I_{Y_{D_2}}}\Bigg] \ge \Theta^{(1)}_{2} \Bigg\rbrace, \nonumber
\end{equation}
When $\Theta^{(1)}_{1} > a_1/ a_2$, then $\mathbb{P}\left(\textrm{DT}^C_{SD_2}\right)=0$, 
otherwise we continue the derivation. We set $\mathcal{G}^{(1)}_{2}= \Theta^{(1)}_{2} /a_2$, then
\begin{equation}
\mathbb{P}\left(\textrm{DT}^C_{SD_2}\right)=\\\mathbb{E}_{I_{X},I_Y}\Bigg[\mathbb{P}\Bigg\lbrace\vert h_{SD_2}\vert^{2}\ge\frac{\mathcal{G}^{(1)}_{1}}{l_{SD_2}}\big[I_{X_{D_2}}+I_{Y_{D_2}}\big], \vert h_{SD_2}\vert^{2}\ge\frac{\mathcal{G}^{(1)}_{2}}{l_{SD_2}}\big[I_{X_{D_2}}+I_{Y_{D_2}}\big]\Bigg\rbrace\Bigg].\nonumber
\end{equation}
Finally, $\mathbb{P}\left(\textrm{DT}^C_{SD_2}\right)$ equals
\begin{equation}
\mathbb{P}\left(\textrm{DT}^C_{SD_2}\right)=\mathcal{L}_{I_{X_{D_2}}}\bigg(\frac{\mathcal{G}^{(1)}_{\mathrm{max}}}{l_{SD_2}}\bigg)\mathcal{L}_{I_{Y_{D_2}}}\bigg(\frac{\mathcal{G}^{(1)}_{\mathrm{max}}}{l_{SD_2}}\bigg),
\end{equation}
where $\mathcal{G}^{(1)}_{\mathrm{max}}=\mathrm{max}(\mathcal{G}^{(1)}_{1},\mathcal{G}^{(1)}_{2})$.\\
To calculate $\mathbb{P}\left(\textrm{RT}^C_{S,D_1,D_2}\right)$, we follow the same steps as in $\mathbb{P}\left(\textrm{RT}^C_{S,D_2,D_1}\right)$.

{
\tiny
\bibliographystyle{ieeetr}
\bibliography{bibnoma}
}
\end{document}